\documentclass[prl,twocolumn,showpacs]{revtex4}
\usepackage{graphics}
\usepackage{epsfig}
\usepackage{graphicx}

\begin{document}
\title{Optimal two-qubit quantum circuits using exchange
interactions}
\author{Heng Fan, Vwani Roychowdhury, Thomas Szkopek}
\affiliation{Department of Electrical Engineering,
University of California Los Angeles, Los Angeles,
CA 90024
}

\date{\today}

\begin{abstract}
We give the optimal decomposition of a universal two-qubit circuit using Heisenberg exchange
interactions and single qubit rotations. Tuning the strength and duration of Heisenberg exchange
allows one to implement $(SWAP)^\alpha $ gates. Our optimal circuit is constructed from only {\em
three} $(SWAP)^\alpha$ gates and {\em six} single-qubit gates. We show that three $(SWAP)^\alpha $
gates are not only sufficient, but necessary. Since six single-qubit gates are already known to be
necessary, our implementation is optimal in gate count.
\end{abstract}

\maketitle

{\it Introduction~}
In several general solid-state quantum computation appoaches\cite{LD,K,VYW,BKL,KBL,DBK}, two-qubit
interactions are generated by a tunable exchange interaction. For example, Heisenberg exchange
between two electron spin qubits results in a $(SWAP)^{\alpha }$ gate, where the exponent $\alpha$
is controlled by adjusting the strength and duration of Heisenberg exchange. Single-qubit rotations
have also been proposed for solid-state computation; notable mechanisms for rotating spin qubits
being g-tensor resonance\cite{KMDG,YJKR} and localized magnetic resonance\cite{LT}. In general, it
is desirable to optimize quantum circuits with respect to the number of physical operations
required, which for most solid-state quantum computation proposals implies that circuits should be
optimized with respect to the number of $(SWAP)^{\alpha }$ operations and single-qubit rotations.

The problem of optimizing quantum circuits for executing general $n$-qubit operations is
computationally intractable. Hence, just as in the classical computation case, one needs to develop
techniques for optimizing only few-qubit circuits and then assemble these circuits together in a
modular fashion. Toward this end, circuit optimization results have mostly dealt with the case
where controlled-NOT (CNOT) gates and single-qubit rotations are the basic building blocks. For
example, for a general two-qubit unitary operation, it has been recently shown that three CNOT
gates and additional single-qubit rotations are sufficient and necessary \cite{VD,VW}. Now, it is
known that one CNOT gate can be realized by two $(SWAP)^{1/2}$ gates and single-qubit unitary
gates\cite{LD}. Hence, six $(SWAP)^{1/2}$ gates are sufficient to implement any two-qubit
operation. The question is whether this strategy of simple substitution is optimal? Or, are
$(SWAP)^{\alpha }$ gates just as efficient as CNOT gates (in terms of gate count) in performing
two-qubit operations? Our answer to the latter is in the affirmative: The {\em $(SWAP)^{\alpha }$
gates and CNOT gates are both equally efficient} at realizing any two-qubit quantum operation (when
measured in terms of number of gates), and that, in order to achieve optimal realizations {\em each
type of circuit requires its own optimization scheme}.

The primary results of our paper are as follows. An arbitrary two-qubit operation can be
implemented using only three $(SWAP)^{\alpha }$ gates and six single-qubit rotations. We augment
this result with a number of lower bounds. First, we show that, by considering entanglement power
alone, a CNOT gate requires at least two $(SWAP)^{\alpha}$ gates. Next, we prove that three
$(SWAP)^{\alpha }$ gates are not only sufficient, but in fact necessary, to implement an arbitrary
two-qubit operation. Our universal two-qubit circuit is optimal in the number of both
$(SWAP)^{\alpha}$ and single-qubit gates.

{\it Heisenberg interaction~} Let's first fix some notation; the four Bell states are written as, $
|\Phi^\pm \rangle =\frac {1}{\sqrt{2}}(|00\rangle \pm |11\rangle ) $, $ |\Psi^\pm \rangle =\frac
{1}{\sqrt{2}}(|01\rangle \pm |10\rangle ) $. The SWAP gate is defined as $SWAP|\psi \rangle |\phi
\rangle =|\phi \rangle |\psi \rangle $. For a ${\cal{C}}^2\otimes {\cal {C}}^2$ system,  it can be
written explicitly as,
\begin{eqnarray}
SWAP&=&|\Phi ^+\rangle \langle \Phi ^+|+|\Phi ^-\rangle \langle \Phi ^-| +|\Psi ^+\rangle \langle
\Psi ^+|
\nonumber \\
&&+e^{i\pi }|\Psi ^-\rangle \langle \Psi ^-|.
\end{eqnarray}
In this paper, our basic two-qubit gate is $(SWAP)^{\alpha }$, it can be written as,
\begin{eqnarray}
(SWAP)^{\alpha }&=&
|\Phi ^+\rangle \langle \Phi ^+|+|\Phi ^-\rangle \langle \Phi ^-|
+|\Psi ^+\rangle \langle \Psi ^+|
\nonumber \\
&&+e^{i\pi \alpha } |\Psi ^-\rangle \langle \Psi ^-|,
\nonumber \\
&=&
\left( \begin{array}{cccc}
1&0&0&0\\
0&\frac {1+e^{i\pi \alpha }}{2}&
\frac {1-e^{i\pi \alpha }}{2}&0\\
0&\frac {1-e^{i\pi \alpha }}{2}&
\frac {1+e^{i\pi \alpha }}{2}&0\\
0&0&0&1\end{array}\right) .
\label{swap}
\end{eqnarray}

The Hamiltonian of the isotropic Heisenberg exchange interaction between electron spins $\vec{S}_1$
and $\vec{S}_2$ is,
\begin{eqnarray}
{\cal {H}}=J(t)\vec{S}_1\cdot \vec{S}_2,
\label{Hamil}
\end{eqnarray}
where $\vec{S}=\{ \sigma _x,\sigma _y,\sigma _z\} $ is a vector of Pauli matrices,
\begin{eqnarray}
\sigma _x=\left( \begin{array}{cc}
0&1\\ 1&0\end{array}\right) ,
\sigma _y=\left( \begin{array}{cc}
0&-i\\ i&0\end{array}\right) ,
\sigma _z=\left( \begin{array}{cc}
1&0\\ 0&-1\end{array}\right) .
\end{eqnarray}
The coupling constant $J(t)$ can in principal be tuned for confined electrons\cite{LD}. The unitary
operator generated by this Hamiltonian is,
\begin{eqnarray}
U_{12}={\rm exp}\left(-\frac{i}{\hbar} \vec{S}_1\cdot \vec{S}_2 \int J(t) dt  \right).
\label{utime}
\end{eqnarray}
By adjusting the integrated coupling $\int J(t) dt$, the unitary operator $U_{12}$ can naturally
realize the gate $(SWAP)^{\alpha }$ where $\alpha = \int J(t) dt / h$. In this paper, we will use
the $(SWAP)^{\alpha }$ gate as the two-qubit gate. It was proposed to use the Heisenberg
interaction {\em alone} to implement quantum computing\cite{BKL,KBL,DBK}. This scheme encodes one
logical qubit as three physical qubits. Additionally, one CNOT gate requires 19 Heisenberg
interactions amongst the six physical qubits. We consider the scheme where {\em both Heisenberg
interaction}, as well as, {\em single-qubit rotations} are available.

{\it CNOT gate requires two $(SWAP)^{\alpha }$ gates~} We know that one CNOT gate can be realized
by two square root of SWAP gates, $\sqrt{SWAP}$. If we use a more general gate, $(SWAP)^{\alpha }$,
can we realize the CNOT by only one $(SWAP)^{\alpha }$ gate and a certain number of single-qubit
rotations? By studying the non-local invariants of the quantum gates, Makhlin showed that the CNOT
gate cannot be constructed by applying the Heisenberg interaction $\cal{H}$ only once, i.e., two
$(SWAP)^{\alpha }$ gates are necessary to construct one CNOT\cite{M}. In this section, we give a
different proof, using entanglement power, to show that the CNOT gate requires at least two
$(SWAP)^{\alpha }$ gates.

The entanglement power of a unitary operator $U\in SU(4)$ is defined as,
\begin{eqnarray}
E_p(U)={\rm Average}_{|\psi _1\rangle \otimes |\psi _2\rangle }
[E(U|\psi _1\rangle \otimes |\psi _2\rangle )],
\end{eqnarray}
where the average is over all product states $|\psi _1\rangle \otimes |\psi _2\rangle \in {\cal
{C}}^2\otimes {\cal {C}}^2$ in uniform distribution, see \cite{ZZF}, and $E$ is the linear entropy
which is also the concurrence\cite{W}. Note that for arbitrary $U_1\otimes U_2\in SU(2)\otimes
SU(2)$, $E_p(U)=E_p(U_1\otimes U_2U)$. So, the entanglement power of $(u_1\otimes
v_1)(SWAP)^{\alpha }(u_2\otimes v_2)$ is actually the entanglement power of $(SWAP)^{\alpha }$.

A simple formula can be used to calculate the entanglement power\cite{ZZF,VW},
\begin{eqnarray}
E_p(U)=\frac {5}{9}-\frac {1}{36}
[\langle U^{\otimes 2}, T_{1,3}U^{\otimes 2}T_{1,3}\rangle
\nonumber \\
+\langle (SWAP\cdot U)^{\otimes 2}, T_{1,3}(SWAP\cdot U)^{\otimes 2}
T_{1,3}\rangle ],
\end{eqnarray}
where $T_{1,3}$ acting on ${\cal {C}}^2\otimes {\cal {C}}^2\otimes
{\cal {C}}^2\otimes
{\cal {C}}^2$ is the transposition operator: $T_{1,3}|a,b,c,d\rangle
=|c,b,a,c\rangle $. By tedious but straight forward calculations we can show that,
\begin{eqnarray}
E_p((SWAP)^{\alpha })=\frac {1}{12}-\frac {1}{12}\cos (2\pi \alpha ).
\label{epower}
\end{eqnarray}
For detailed calculations, see Appendix A in Ref.\cite{FRS}.
When $\alpha =1/2$, $(SWAP)^{\alpha }$ has a maximum entanglement power of $1/6$. Since the
entanglement power of CNOT is $2/9$ \cite{VW}, which is strictly larger than $1/6$,
one $(SWAP)^{\alpha }$ operator with
the help of single-qubit gates is not sufficient to realize the CNOT. Hence, at least two
$(SWAP)^{\alpha }$ gates are necessary to realize a general $SU(4)$ operator.

{\it General two-qubit operation~}
Kraus and Cirac \cite{KC} gave the following result, (see also \cite{KBG}): an arbitrary unitary
transformation $U\in SU(4)$ has the decomposition,
\begin{eqnarray} \label{eqn:decomp}
U=(u_4'\otimes v_4')e^{-iH}(u_1\otimes v_1),
\end{eqnarray}
where
$u_1, v_1, u_4', v_4'\in SU(2)$, and
\begin{eqnarray}
H\equiv h_x\sigma _x\otimes \sigma _x
+h_y\sigma _y\otimes \sigma _y
+ h_z\sigma _z\otimes \sigma _z,
\label{H}
\end{eqnarray}
where $\pi /4\ge h_x\ge h_y\ge h_z\ge 0$. Then $H$ in (\ref{H}) can be written as,
\begin{eqnarray}
H&=&\lambda_{00} |\Phi^+ \rangle \langle \Phi^+ | + \lambda_{01} |\Psi^+ \rangle \langle \Psi^+ | +
\lambda_{10} |\Phi^- \rangle \langle \Phi^-| \nonumber \\ &&+ \lambda_{11} |\Psi^- \rangle \langle
\Psi^-|, \label{H1}
\end{eqnarray}
with,
\begin{eqnarray}
\lambda _{00}=h_x-h_y+h_z, ~~
\lambda _{01}=h_x+h_y-h_z,\nonumber \\
\lambda _{10}=-h_x+h_y+h_z, ~~
\lambda _{11}=-h_x-h_y-h_z.
\label{kcresult}
\end{eqnarray}
The diagonal form of $H$ thus gives,
\begin{eqnarray}
e^{-iH}=e^{-i\lambda_{00}} |\Phi^+ \rangle \langle \Phi^+ | + e^{-i\lambda_{01}} |\Psi^+ \rangle
\langle \Psi^+ | \nonumber \\ + e^{-i\lambda_{10}} |\Phi^- \rangle \langle \Phi^-| +
e^{-i\lambda_{11}} |\Psi^- \rangle \langle \Psi^-|. \label{diag}
\end{eqnarray}

Vidal and Dawson\cite{VD}, Vatan and Williams \cite{VW} have shown that the operator $e^{-iH}$ can
be realized by only three CNOT gates and some single-qubit rotation gates. Thus an arbitrary $U\in
SU(4)$ can be realized by three CNOT gates and additional single-qubit gates, see also the
appendix B in Ref.\cite{FRS}.

{\it Arbitrary two-qubit unitary operations require only three $(SWAP)^{\alpha}$
gates and six single-qubit gates~}

Recall that a CNOT gate can be realized by two $(SWAP)^{1/2}$ gates and a few extra single-qubit
operations. We know the optimal circuit for a general $U\in SU(4)$ needs three CNOT gates, so six
$(SWAP)^{\alpha }$ gates are needed if we simply substitute SWAP circuits for CNOT gates.
Our aim is to find a
circuit to realize $U\in SU(4)$ optimal in the number of $(SWAP)^{\alpha }$ gates.

From the result of Kraus and Cirac (Eq. \ref{diag}) \cite{KC}, we need to create arbitrary phases
on four Bell states by $(SWAP)^{\alpha }$ gates. But we already know that $(SWAP)^{\alpha }$
applies a phase to the Bell state $|\Psi ^-\rangle $ while leaving the other three Bell states
invariant. Also, by Pauli rotations on one particle of the bipartite state, we can transform the
Bell states amongst each other. Thus, it is straightforward to create four independent phases on
the Bell states. We can rewrite the operator $\exp(-iH)$ as,
\begin{eqnarray}
e^{-iH}&=&e^{i(h_z-h_x-h_y)}
(|\Psi ^+\rangle \langle \Psi ^+|
+e^{2i(h_x+h_y)}|\Psi ^-\rangle \langle \Psi ^-|
\nonumber \\
\hspace*{-7mm}&+&e^{2i(h_y-h_z)}|\Phi ^+\rangle \langle \Phi ^+|
+e^{2i(h_x-h_z)}|\Phi ^-\rangle \langle \Phi ^-|).
\end{eqnarray}
This operator can be constructed by
$(SWAP)^{\alpha }$ operators as
\begin{eqnarray}
&&e^{-iH}=e^{i(h_z-h_x-h_y)}\times
\nonumber \\
&&\left[
(I\otimes \sigma _x\sigma _z)
(SWAP)^{2(h_y-h_z)/\pi }(I\otimes \sigma _z\sigma _x)\right]\times
\nonumber \\
&&\left[
(I\otimes \sigma _x)(SWAP)^{2(h_x-h_z)/\pi }(I\otimes \sigma _x)
\right] \times
\nonumber \\
&&\left[
(SWAP)^{2(h_x+h_y)/\pi }\right]
\end{eqnarray}
This circuit just involves three $(SWAP)^{\alpha }$ operators and single-qubit gates which are
Pauli matrices. So as a whole, we can construct any $U\in SU(4)$ by only three SWAP gates and
single-qubit rotations. Note that besides Pauli matrices, general single-qubit rotations are also
necessary to transform $e^{-iH}$ to $U$.

Up to an overall phase, we can rewrite $e^{-iH}$,
\begin{eqnarray}
e^{-iH}&=&(\sigma _z\otimes \sigma _x)(SWAP)^{\gamma }(\sigma _z\otimes I)
\nonumber \\
&&
(SWAP)^{\beta }(I\otimes \sigma _x)(SWAP)^{\alpha },
\label{optcircuit}
\end{eqnarray}
where $\alpha =2(h_x+h_y)/\pi ,\beta =2(h_x-h_z)/\pi , \gamma =2(h_y-h_z)/\pi .$ The corresponding
circuit for $U$ is illustrated in Fig. \ref{eqn:decomp}.

\begin{figure}[ht]
\includegraphics[width=8cm]{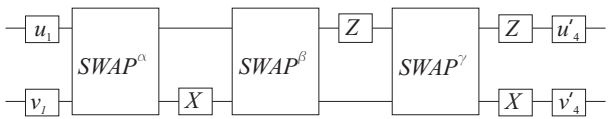}
\caption{Circuit for arbitrary unitary transformation $U \in SU(4)$ as decomposed in Eqs.
(\ref{eqn:decomp}) and (\ref{optcircuit}). Three SWAP gates, and 6 local unitaries (upon combination of
$u'_4$ and $v'_4$ with $Z$ and $X$, respectively) are required. }
\end{figure}

In any implementation, single-qubit rotations as well as two-qubit operations will require physical
resources such as time and hardware. Hence, it is helpful to consider the number of single-qubit
gates involved in each circuit as well. In the circuit of Vidal and Dawson\cite{VD}, eight
single-qubit gates are used to construct the general $U\in SU(4)$, while in our circuit , six
single-qubit gates are used. It is known that this is the least possible number of single-qubit
rotations\cite{ZVSW}. If we assume that each single-qubit rotation is as expensive as a two-qubit
operation, then our circuit is potentially {\it cheaper}.

We next show that our circuit is optimal in the number of $(SWAP)^{\alpha }$ gates
used. i.e., three $(SWAP)^{\alpha }$ gates
are \emph{necessary} to construct a general circuit.  Let's write out a general unitary operator
which contains just two $(SWAP)^{\alpha }$ gates,
\begin{eqnarray}
U &=& ( U_1 \otimes V_1 ) (SWAP)^{\alpha} ( U_2 \otimes V_2 )
\nonumber \\
&& (SWAP)^{\beta} ( U_3 \otimes V_3 ),
\end{eqnarray}
where $U_j,V_j, j=1,2,3$ are single-qubit operations.

One can group single-qubit unitaries about the $(SWAP)^{\alpha }$ operators as follows,
\begin{eqnarray}
U &=& ( U_1 \otimes V_1 ) (SWAP)^{\alpha} (U_1^{\dagger }
\otimes V_1^{\dagger })
\nonumber \\
&&(\tilde {U}_2\otimes \tilde {V}_2)(SWAP)^{\beta }(\tilde {U}_2^{\dagger }
\otimes \tilde {V}_2^{\dagger })(\tilde {U}_3\otimes \tilde {V}_3),
\end{eqnarray}
where $ \tilde {U}_2=U_1 U_2$, $\tilde {V}_2=V_1V_2$, $\tilde {U}_3=U_1 U_2 U_3$, and $\tilde
{V}_3=V_1V_2V_3$. Here we notice that operators $( U_1 \otimes V_1 ) (SWAP)^{\alpha} (U_1^{\dagger
} \otimes V_1^{\dagger })$ and $(\tilde {U}_2\otimes \tilde {V}_2)(SWAP)^{\beta }(\tilde
{U}_2^{\dagger } \otimes \tilde {V}_2^{\dagger })$ are just SWAP gates in some different basis. So
we can write this relation as,
\begin{eqnarray}
U=(SWAP)^{\alpha }(\widetilde{SWAP})^{\beta }(u\otimes v).
\end{eqnarray}
A single $(SWAP)^{\alpha }$ gate can create one phase in one maximally entangled state, with the
orthogonal three dimensional space left invariant. So for two $(SWAP)^{\alpha }$ gates, there exist
two maximally entangled orthogonal states, $|\chi_1\rangle$ and $|\chi_2\rangle$, simultaneously
orthogonal to $|\Psi ^-\rangle $ and $|\widetilde {\Psi }^-\rangle $. >From the symmetry of the
$(SWAP)^{\alpha }$ operation, it follows that for every two-$(SWAP)^{\alpha }$-gate circuit there
exists at least two orthogonal and maximally entangled states such that they cannot be assigned a
relative phase by the circuit. That is, the unitary operator corresponding to the
two-$(SWAP)^{\alpha }$-gate circuit must satisfy $U|\chi_j\rangle =(u\otimes v) |\chi_j\rangle
,j=1,2$. Note that local unitary operations cannot add a relative phase to two maximally entangled
states. However, since a general two-qubit operation can assign independent phases to three
maximally entangled states, one can find $U\in SU(4)$ such that it will never satisfy the preceding
constraint. Hence, two $(SWAP)^{\alpha }$ gates and single-qubit rotations are not sufficient to
construct an arbitrary $U\in SU(4)$.

{\it Spin-based quantum computation by SWAP circuit~}

In this section, we will estimate the real time required to implement our $(SWAP)^{\alpha }$
circuit. We consider both the GaAs and Si semiconductor systems, for review see Ref.\cite{H}. A
reasonable effective resonant magnetic field is $B=1\mathrm{mT}$, giving a Rabi frequency for an
electron in GaAs (Si) of approximately 6.2MHz ( 28MHz ), so that a single qubit $\pi$ rotation
requires approximately 80ns (18ns). The effective resonant magnetic field could be generated by
localized magnetic excitation \cite{LT} or through g-tensor modulation \cite{YJKR,KMDG}. A SWAP
gate needs about 50ps for $J \approx 0.1$meV (\ref{Hamil}), and hence 50ps is the maximum time
required for performing a $(SWAP)^{\alpha}$ gate.

From the point of view of operating time (and hence qubit storage errors), reducing the number of
single qubit gates becomes the overwhelming consideration in designing a circuit. Our circuit is
formed by at most three $(SWAP)^{\alpha}$ gates and six single qubit rotations. Considering that
single qubit rotations can in principle be performed on individual qubits simultaneously, the total
time to implement a general $(SWAP)^{\alpha }$ circuit is at most the time required for three
single qubit rotations. The optimal CNOT circuit given by Ref.\cite{VD} contains three CNOT gates
and eight single qubit rotations. If we assume that a CNOT gate takes almost the same time as a
SWAP gate, we find that the time to implement the CNOT circuit is also at most the time to
implement three single qubit rotations (distinct from the single qubit rotations of the
$(SWAP)^{\alpha }$ circuit). In practice, one would like to implement specific instances of $U \in
SU(4)$, so that gate counts and timings depend on the target operation $U$. The self-evident
example is that it is best to implement $U=\mathrm{CNOT}$ with a CNOT circuit, and similarly
$U=\mathrm{SWAP}$ with a $(SWAP)^{\alpha}$ circuit. The essential point is that given the physical
means to implement single qubit rotations and $(SWAP)^{\alpha}$, we can arrive at the most
efficient decomposition (in terms of operating time and gate count) of a target $U \in SU(4)$ that
might be specified using a network of CNOT's, controlled-phase gates or other logically convenient
gates. This optimization is essential for minimizing both storage and gate errors introduced by
performing the operation $U$.

The entanglement power of $(SWAP)^{\alpha}$ gate is strictly less than the CNOT gate, but
$(SWAP)^{\alpha}$ gate is as efficient as CNOT gate (in terms of gate count) in performing
two-qubit operations. From the point of view of operating time, we also showed that both
$(SWAP)^{\alpha}$ circuit and CNOT circuit is the same for a general two-qubit unitary operation.
So the optimal $(SWAP)^{\alpha}$ circuit is as good as the optimal CNOT circuit if the complexity
of physical implementation of two-qubit gates (CNOT gate and $(SWAP)^{\alpha}$ gate) and
single-qubit gates are the same. However if we directly replace a CNOT gate by two $(SWAP)^{1/2}$
gates and some single qubit gates to implement a general two-qubit operation, we need six
$(SWAP)^{1/2}$ gates and the operating time will be at least doubled compared with the optimal
$(SWAP)^{\alpha}$ circuit. In this sense the optimizition of $(SWAP)^{\alpha}$ circuit is
necessary, the advantages of the optimization are: first, the number of two-qubit gates is reduced
by half; second,the operating time is reduced to one half to one third due to the reduction of the
number of single qubit gates.

Our circuit realizes the three free parameters in Eq.(\ref{kcresult}) by adjusting the $\alpha $ in
$(SWAP)^{\alpha }$ gate. So the SWAP circuit generally involves three different $(SWAP)^{\alpha }$
gates with fixed single qubit rotations. This is in contrast with the CNOT circuit, in which the
two-qubit operation, the CNOT gate, is fixed, and the single qubit rotations are tuned according to
the free parameters in Eq.(\ref{kcresult}). This is a trade-off in designing a circuit, which do we
prefer: tuning $(SWAP)^{\alpha}$ gates or tuning single qubit rotations? As is generally accepted,
in solid state with Heisenberg exchange interaction, the $(SWAP)^{\alpha }$ gate can be realized by
simply controlling the interaction time as presented in (\ref{utime}). An arbitrary single qubit
rotation is potentially more difficult to realize and may take up to 100ns \cite{BLD} time to
process. For example, one can apply magnetic time varying magnetic fields. Since the single qubit
rotations are fixed in the SWAP circuit of (\ref{optcircuit}), (the single qubit rotations at the
front and back of the circuit still need be adjusted according to the specified operations), the
directions, gradients, pulse time etc. of the magnetic field can be fixed. In this sense, for
spin-based quantum computation, tuning the $(SWAP)^{\alpha }$ gate in our circuit could be
advantageous compared to using a fixed $(SWAP)^{\alpha }$ circuit, such as $(SWAP)^{1/2}$, and
tuning the single qubit rotations.

{\it Summary and Discussion ~}

When it comes to solid state implementations, the exchange interaction has emerged as the primary
mechanism for constructing non-local quantum gates, and the $(SWAP)^{\alpha }$ gate is the cheapest
and the most natural two-qubit gate that can be realized using this technology. We have shown that
simply replacing individual CNOT gates with its SWAP circuit is not an efficient implementation
technique for exchange-interaction based quantum computing systems. We have presented an alternate
optimization technique and have derived the optimal circuit for an arbitary two-qubit unitary
operator using SWAP gates and single-qubit rotations.

\noindent {\textbf{Acknowledgement :} This work was sponsored in part by the U.S. Army Research
Office/DARPA under contract/grant number DAAD 19-00-1-0172, and in part by the NSF under contract
number CCF-0432296.

{\it Appendix A: Entanglement power of $(SWAP)^{\alpha }$ gate~}
As presented in section III,
The entanglement power of a unitary operator
$U\in {\cal {C}}^2\otimes {\cal {C}}^2$ is defined as,
\begin{eqnarray}
E_p(U)={\rm Average}_{|\psi _1\rangle \otimes |\psi _2\rangle }
[E(U|\psi _1\rangle \otimes |\psi _2\rangle )],
\end{eqnarray}
where the average is over all product states $|\psi _1\rangle \otimes |\psi _2\rangle \in {\cal
{C}}^2\otimes {\cal {C}}^2$ in uniform distribution, see \cite{ZZF}.
Zanardi $et~al$\cite{ZZF} showed the following result:
\begin{eqnarray}
E_p(U)&=&1-(C_{2})^2\sum _{k=0,1}I_k(U),
\nonumber \\
I_k(U)&=&{\rm tr}T_{1+k,3+k}
+\langle U^{\otimes 2}(T_{1+k,3+k})U^{\dagger \otimes 2}, T_{1,3}\rangle ,
\nonumber
\end{eqnarray}
where the scalar product $\langle A,B\rangle :={\rm tr}(A^{\dagger }B)$,
$T_{1,3}$ and $T_{2,4}$ are the transposition operators:
$T_{1,3}|a,b,c,d\rangle =|c,b,a,d\rangle $,
$T_{2,4}|a,b,c,d\rangle =|a,d,c,b\rangle $.
For case $U\in {\cal {C}}^2\otimes {\cal {C}}^2$, we know
that $C_2=6$, and ${\rm tr}T_{1,3}={\rm tr}T_{2,4}=8$.
With the help of $T_{2,4}=(SWAP\otimes SWAP)T_{1,3}(SWAP\otimes SWAP)$,
we know that for $U\in {\cal {C}}^2\otimes {\cal {C}}^2$, the
entanglement power of $U$ is
\begin{eqnarray}
E_p(U)=\frac {5}{9}-\frac {1}{36}
[\langle U^{\otimes 2}, T_{1,3}U^{\otimes 2}T_{1,3}\rangle
\nonumber \\
+\langle (SWAP\cdot U)^{\otimes 2}, T_{1,3}(SWAP\cdot U)^{\otimes 2}
T_{1,3}\rangle ]
\end{eqnarray}
as presented in section III and in Ref.\cite{VW}.

Our aim is to find the entanglement power of $(SWAP)^{\alpha }$.
Subsititute $U$ by $(SWAP)^{\alpha }$ in (\ref{swap}), we can find
the second term in (\ref{epower}) can be calculated as
\begin{eqnarray}
&&
\langle U^{\otimes 2}, T_{1,3}U^{\otimes 2}T_{1,3}\rangle
\nonumber \\
&=&\frac {17}{2}+3(e^{i\pi \alpha }+e^{-i\pi \alpha })
+\frac {3}{4}(e^{2i\pi \alpha }+e^{-2i\pi \alpha }).
\end{eqnarray}
We can find that $SWAP\cdot (SWAP)^{\alpha }$ is similar to
$(SWAP)^{\alpha }$, we just need to replace $e^{i\pi \alpha }$
by $-e^{i\pi \alpha }$ in (\ref{swap}). Thus the third term
in (\ref{epower}) is
\begin{eqnarray}
&&\langle (SWAP\cdot U)^{\otimes 2}, T_{1,3}(SWAP\cdot U)^{\otimes 2}
T_{1,3}\rangle
\nonumber \\
&=&\frac {17}{2}-3(e^{i\pi \alpha }+e^{-i\pi \alpha })
+\frac {3}{4}(e^{2i\pi \alpha }+e^{-2i\pi \alpha }).
\end{eqnarray}
With all of these results we know the entanglement power of
$(SWAP)^{\alpha }$ is
\begin{eqnarray}
E_p((SWAP)^{\alpha })=\frac {1}{12}
-\frac {1}{12}\cos (2\pi \alpha ).
\end{eqnarray}
We can find that the entanglement power of
$(SWAP)^{\alpha }$ is greater than zero except
$\alpha =0$ and $\alpha =1$ which corresponding to
the identity and the full SWAP
respectively.
The $\sqrt{SWAP}$ has the maximal
entanglement power but it is less than the CNOT gate.

{\it Appendix B, Optimal circuit by CNOT gates and single
qubit rotations~}
We review here Vidal and Dawson's CNOT circuit. We know that the general unitary operator can be
simplified by single-qubit rotations to an operator $e^{-iH}$ which can create arbitrary phases on
four Bell states. Using CNOT gates, we need to find a circuit which can create three independent
phases on four Bell states; note that an overall phase is not important here. It is well known that
by CNOT and Hadamard transformation, $(W\otimes I)CNOT$, we can transform the Bell basis to the
computational basis $|\Phi ^+\rangle \rightarrow |00\rangle $, $|\Phi ^-\rangle \rightarrow
|10\rangle $, $|\Psi ^+\rangle \rightarrow |01\rangle $, $|\Psi ^-\rangle \rightarrow |11\rangle $.
The Hadamard transformation is defined as $W=\left( \begin{array}{cc}1&1\\1&-1\end{array}\right) /2
$. The inverse operator CNOT$(W\otimes I)$ transfers the computational basis back to Bell basis.
Since the operator ${\rm exp}(-i\zeta \sigma _z)$ applies phases $|0\rangle \rightarrow e^{-i\zeta
}|0\rangle $, $|1\rangle \rightarrow e^{i\zeta }|1\rangle $, we can construct a circuit,
\begin{eqnarray}
CNOT(W\otimes I)(e^{-i\zeta _2\sigma _z}\otimes e^{-i\xi _2\sigma _z})
CNOT
\nonumber \\
(e^{-i\zeta _1\sigma _z}\otimes e^{-i\xi _1\sigma _z})
(W\otimes I)CNOT.
\label{circuit0}
\end{eqnarray}
By this circuit, we can apply arbitrary phases on the four Bell states,
\begin{eqnarray}
|\Phi^+\rangle & \rightarrow & e^{-i(+\zeta _1+\xi _1+\zeta _2+\xi _2)} |\Phi^+\rangle,
\nonumber \\
|\Psi^+\rangle & \rightarrow & e^{-i(+\zeta _1-\xi _1+\zeta _2-\xi _2)} |\Psi^+\rangle,
\nonumber \\
|\Phi^-\rangle & \rightarrow & e^{-i(-\zeta _1-\xi _1-\zeta _2+\xi _2)} |\Phi^-\rangle,
\nonumber \\
|\Psi^-\rangle & \rightarrow & e^{-i(-\zeta _1+\xi _1-\zeta _2-\xi _2)} |\Psi^-\rangle.
\end{eqnarray}
The parameters $\zeta _1,\zeta _2,\xi _1, \xi _2$ can be chosen so as to reproduce $\exp(-iH)$ in
Eq. \ref{diag}, thus recovering the result given by Vidal and Dawson\cite{VD}. Here we have
provided a more intuitive understanding of their results.

The circuit (\ref{circuit0}) is optimal in the sense that the number of CNOT gates in it is
minimized. Vidal and Dawson\cite{VD}, Vatan and Williams \cite{VW} gave two different proofs to
show that at least three CNOT gates are necessary. Here we provide another proof which combines
both. In Ref.\cite{VD}, it is shown that the most general circuit involving two CNOT gates can be
simplified to $CNOT(e^{-i\zeta \sigma _x}\otimes e^{-i\xi \sigma _z})CNOT$. In Ref.\cite{VW}, it is
shown that the simplified circuit is $CZ(e^{-i\zeta \sigma _x}\otimes e^{-i\xi \sigma _x})CZ$,
where CZ is the controlled-phase gate. Our observation here is that these two simplified circuits
only have two free parameters. With additional single-qubit rotations applied before and after the
CNOT's, we can have at most $12+2=14$ free parameters. But we need at least 15 since the two-qubit
operator acts in a 4-dimensional Hilbert space, thus at least three CNOT gates are necessary. This
is a simple proof that the circuit of Eq. \ref{circuit0} is optimal.


\begin{thebibliography}{99}
\bibitem{LD}D.Loss and D.DiVincenzo, Phys.Rev.A {\bf 57}, 120 (1998).
\bibitem{K}B.E.Kane, Nature {\bf 393}, 133 (1998).
\bibitem{VYW}R.Vrijen, E.Yablonovitch, K.L.Wang, H.W.Jiang,
A.Balandin, V.Roychowdhury, T.Mor, D.P.DiVincenzo,
Phys.Rev.A{\bf 62}, 012306 (2000).
\bibitem{BKL}D.Bacon, J.Kempe, D.A.Lidar, and K.B.Whaley,
Phys. Rev. Lett. {\bf 85}, 1758 (2000).
\bibitem{KBL}J.Kempe, D.Bacon, D.A.Lidar, and K.B.Whaley,
Phys. Rev. A {\bf 63}, 042307 (2001).
\bibitem{DBK}D.P.DiVincenzo, D.Bacon, J.Kempe, G.Burkard, and
K.B.Whaley, Nature {\bf 408}, 339 (2000).
\bibitem{YJKR}E.Yablonovitch, H.W.Jiang, H.Kosaka, H.D.Robinson,
D.S.Rao, and T.Szkopek, Proc. IEEE {\bf 91}, 761 (2003).
\bibitem{KMDG}Y.Kato, R.C.Meyers, D.C.Driscoll, A.C.Gossard,
J.Levy and D.D.Awschalom, Science {\bf
299}, 1201, (2003).
\bibitem{LT}D.A.Lidar,J.H.Thywissen,J.Appl.Phys.{\bf 96},754(2004).
\bibitem{VD}G.Vidal,C.M.Dawson,Phys.Rev.A{\bf 69},010301(R)(2004).
\bibitem{VW}F.Vatan and C.Williams, Phys.Rev.A {\bf 69}, 032315 (2004).
\bibitem{M}Y.Makhlin, Quant. Info. Proc. {\bf 1}, 243 (2002).
\bibitem{ZZF}P.Zanardi, C.Zalka, L.Faoro, Phys.Rev.A {\bf 62},
30301(R), (2000).
\bibitem{W}W.K.Wootters, Phys.Rev.Lett. {\bf 80}, 2245 (1998).
\bibitem{KC}B.Kraus and J.I.Cirac, Phys.Rev.A 63, 062309 (2001).
\bibitem{KBG}N.Khaneja, R.Brockett and S.J.Glaser,
Phys.Rev.A{\bf 63}, 032308 (2001).
\bibitem{ZVSW}J.Zhang, J.Vala, S.Sastry, and B.Whaley,
Phys. Rev. Lett. {\bf 93}, 020502 (2004).
\bibitem{FRS}H.Fan, V.Roychowdhury, T.Szkopek, quant-ph/0410001.
\bibitem{H}X.D.Hu,
cond-mat/0411012.
\bibitem{BLD}G.Burkard, D.Loss, and D.P.DiVincenzo, Phys.
Rev. A59, 2070(1999).
\end{thebibliography}
\end{document}